\documentclass[french,english]{revtex4-1}
\usepackage[T1]{fontenc}
\usepackage[latin9]{inputenc}
\usepackage{color}
\usepackage{amsmath}
\usepackage{amssymb}
\usepackage{wasysym}
\usepackage{esint}

\makeatletter
\usepackage[colorlinks = true,
            linkcolor = red,
            urlcolor  = blue,
            citecolor = blue,
            anchorcolor = blue]{hyperref}

\makeatother

\usepackage{babel}
\makeatletter
\addto\extrasfrench{%
   \providecommand{\fg}{\ifdim\lastskip>\z@\unskip\fi~\frqq}%
}

\makeatother
\begin{document}

\title[\textbf{On the 3D Pauli Eq in NCPS}]{\texttt{ON THE THREE-DIMENSIONAL PAULI EQUATION IN NONCOMMUTATIVE
PHASE-SPACE} }

\author{{\normalsize{}Ilyas Haouam}}

\email{ilyashaouam@live.fr }

\address{Laboratoire de Physique Mathématique et de Physique Subatomique (LPMPS),
Université Frères Mentouri, Constantine 25000, Algeria}
\begin{abstract}
In this paper, we obtained the three-dimensional Pauli equation for
a spin-1/2 particle in the presence of an electromagnetic field in
noncommutative phase-space, as well the corresponding deformed continuity
equation, where the cases of a constant and non-constant magnetic
field are considered. Due to the absence of the current magnetization
term in the deformed continuity equation as expected, we had to extract
it from the noncommutative Pauli equation itself without modifying
the continuity equation. It is shown that the non-constant magnetic
field lifts the order of the noncommutativity parameter in both the
Pauli equation and the corresponding continuity equation. However,
we successfully examined the effect of the noncommutativity on the
current density and the magnetization current. By using a classical
treatment, we derived the semi-classical noncommutative partition
function of the three-dimensional Pauli system of the one-particle
and N-particle systems. Then, we employed it for calculating the corresponding
Helmholtz free energy followed by the magnetization and the magnetic
susceptibility of electrons in both commutative and noncommutative
phase-spaces. Knowing that with both the three-dimensional Bopp-Shift
transformation and the Moyal-Weyl product, we introduced the phase-space
noncommutativity in the problems in question. 

$\phantom{}$

\textbf{\normalsize{}Keywords:} 3-D noncommutative phase-space; Pauli
equation; deformed continuity equation; current magnetization; semi-classical
partition function; magnetic susceptibility 
\end{abstract}

\keywords{3-D noncommutative phase-space, Pauli equation, deformed continuity
equation, current magnetization, semi-classical partition function,
magnetic susceptibility \tableofcontents{}}

\maketitle

\section{Introduction}

It is known that the relativistic wave equation describing the fermions
with spin-1/2 is the Dirac equation; on the other hand, the non-relativistic
wave equation describing them, namely the Pauli equation, which is
a topic of great interest in physics \cite{key-1,key-2,key-3,key-4}.
It is relative to the explanation of many experimental results, and
its probability current density changed to including an additional
spin-dependent term recognized as the spin current \cite{key-5,key-6,key-7}.
Pauli equation shown \cite{key-8,key-9,key-10,key-11,key-12} as the
non-relativistic limit of Dirac equation. Knowing that historically
at first time Pauli in 1927 \cite{key-13} presented his known spin
matrices in modifying the non-relativistic Schrödinger equation to
account for Goudsmit-Uhlenbeck\textquoteright s hypothesis (1925)
\cite{key-14,key-15}. Therefore, he applied an ansatz for adding
a phenomenological term to the ordinary non-relativistic Hamiltonian
in the presence of an electromagnetic field, the interaction energy
of a magnetic field and electronic magnetic moment relative to the
intrinsic spin angular momentum of the electron. Describing this spin
angular momentum through the spin matrices requires replacing the
complex scalar wave function by a two-component spinor wave function
in the wave equation. Since then, the study of the Pauli equation
became a matter of considerable attention.

In 1928 when Dirac presented his relativistic free wave equation in
addition to the minimal coupling replacement to include electromagnetic
interactions \cite{key-16}, he showed that his equation contained
a term involving the electron magnetic moment interacting with a magnetic
field, which was the same one inserted by hand in Pauli\textquoteright s
equation. After that, it became common to account electron spin as
a relativistic phenomenon, and the corresponding spin-1/2 term could
be inserted into the spin-0 non-relativistic Schrodinger equation
as will be discussed in the following to see how this is possible.
However, motivated by attempts to understand string theory and describe
quantum gravitation using noncommutative geometry and by trying to
have drawn considerable attention to the phenomenological implications,
we focus here on studying the problem of a non-relativistic spin-1/2
particle in the presence of an electromagnetic field within 3-dimensional
noncommutative phase-space.

As a mathematical theory, noncommutative geometry is by now well established,
although at first, its progress has been narrowly restricted to some
branches of physics such as quantum mechanics. However, recently,
the noncommutative geometry has become a topic of great interest \textcolor{red}{\cite{key-17,key-18,key-19,key-20,key-21,key-22,key-23}}.
It has been finding applications in many sectors of physics and rapidly
has become involved in them, continued to promote fruitful ideas and
the search for a better understanding. Such as in the quantum gravity
\cite{key-24}; the standard model of fundamental interactions \cite{key-25};
as well in the string theory \cite{key-26}; and its implication in
Hopf algebras \cite{key-27} gives the Connes\textendash Kreimer Hopf
algebras \cite{key-28,key-29,key-30} etc. There are many papers devoted
to the study such various aspects especially in quantum field theory
\cite{key-31,key-32,key-33} and quantum mechanics \cite{key-34,key-35,key-36}. 

This paper is organized as follows. In section 2, we present an analysis
review of noncommutative geometry, in particular both the three-dimensional
Bopp-Shift transformation and the Moyal-Weyl product. In section 3,
we investigate the three-dimensional Pauli equation in the presence
of an electromagnetic field and the corresponding continuity equation.
Besides, we derived the current magnetization term in the deformed
continuity equation. Section 4 is devoted to calculating the semi-classical
noncommutative partition function of the Pauli system of the one-particle
and N-particle systems. Consequently, we obtain the corresponding
magnetization and the magnetic susceptibility through the Helmholtz
free energy, all in both commutative and noncommutative phase-spaces
and within a classical limit. Therefore, concluding with some remarks.

\section{Review of noncommutative algebra}

Firstly, we present the most essential formulas of noncommutative
algebra \cite{key-36}. It is well known that at very tiny scales
such as the string scale, the position coordinates do not commute
with each other, neither do the momenta.

Let us accept in a d-dimensional noncommutative phase-space the operators
of coordinates and momenta $x_{i}^{nc}$ and $p_{i}^{nc}$, respectively.
The noncommutative formulation of quantum mechanics corresponds to
the following Heisenberg-like commutation relations 
\begin{equation}
\begin{array}{cccc}
\left[x_{\mu}^{nc},x_{\nu}^{nc}\right]=i\Theta_{\mu\nu}, & \left[p_{\mu}^{nc},p_{\nu}^{nc}\right]=i\eta_{\mu\nu}, & \left[x_{\mu}^{nc},p_{\nu}^{nc}\right]=i\tilde{\hbar}\delta_{\mu\nu} & ,\:(\mu,\nu=1,..d)\end{array},\label{eq:1}
\end{equation}
the effective Planck constant is the deformed Planck constant, which
is given by
\begin{equation}
\tilde{\hbar}=\alpha\beta\hbar+\frac{\text{Tr}[\Theta\eta]}{4\alpha\beta\hbar},\label{eq:2}
\end{equation}
where $\frac{\text{Tr}[\Theta\eta]}{4\alpha\beta\hbar}\ll1$ is the
condition of consistency in quantum mechanics. $\Theta_{\mu\nu}$,
$\eta_{\mu\nu}$ are constant antisymmetric $d\times d$ matrices
and $\delta_{\mu\nu}$ is the identity matrix. 

It is shown that $x_{i}^{nc}$ and $p_{i}^{nc}$ can be represented
in terms of coordinates $x_{i}$ and momenta $p_{j}$ in usual quantum
mechanics through the so-called generalized Bopp-shift as follows
\cite{key-34} 
\begin{equation}
\begin{array}{cccc}
x_{\mu}^{nc} & = & \alpha x_{\mu}-\frac{1}{2\alpha\hbar}\Theta_{\mu\nu}p_{\nu},\:\text{ and } & p_{\mu}^{nc}=\beta p_{\mu}+\frac{1}{2\beta\hbar}\eta_{\mu\nu}x_{\nu}\end{array},\label{eq:3}
\end{equation}
with $\alpha=1-\frac{\Theta\eta}{8\hbar^{2}}$ and $\beta=\frac{1}{\alpha}$
are scaling constants.

To the 1rst order of $\Theta$ and $\eta$, in the calculations we
take $\alpha=\beta=1$, so the Equations (\ref{eq:3}, \ref{eq:2})
become
\begin{equation}
\begin{array}{ccc}
x_{\mu}^{nc} & =x_{\mu}-\frac{1}{2\hbar}\Theta_{\mu\nu}p_{\nu},\; & p_{\mu}^{nc}=p_{\mu}+\frac{1}{2\hbar}\eta_{\mu\nu}x_{\nu}\end{array},\:\text{ and }\tilde{\hbar}=\hbar+\frac{\text{Tr}[\Theta\eta]}{4\hbar}.\label{eq:4}
\end{equation}

If the system in which we study the effects of noncommutativity is
three-dimensional, we limit ourselves to the following noncommutative
algebra
\begin{equation}
\begin{array}{cccc}
\left[x_{j}^{nc},x_{k}^{nc}\right]=i\frac{1}{2}\epsilon_{jkl}\Theta_{l}, & \left[p_{j}^{nc},p_{k}^{nc}\right]=i\frac{1}{2}\epsilon_{jkl}\eta_{l}, & \left[x_{j}^{nc},p_{k}^{nc}\right]=i\left(\hbar+\frac{\Theta\eta}{4\hbar}\right)\delta_{jk} & ,\:(j,k,l=1,2,3)\end{array},\label{eq:5}
\end{equation}
$\Theta_{l}=(0,0,\Theta)$, $\eta_{l}=(0,0,\eta)$ are the real-valued
noncommutative parameters with the dimension of $length{}^{2}$, $momentum{}^{2}$
respectively, they are assumed to be extremely small. And $\epsilon_{jkl}$
is the Levi-Civita permutation tensor. Therefor, we have 
\begin{equation}
x_{i}^{nc}=x_{i}-\frac{1}{4\hbar}\epsilon_{ijk}\Theta_{k}p_{j}:\begin{cases}
\begin{array}{ccc}
x^{nc} & = & x-\frac{1}{4\hbar}\Theta p_{y}\\
y^{nc} & = & y+\frac{1}{4\hbar}\Theta p_{x}\\
z^{nc} & = & z
\end{array}\end{cases},\;p_{i}^{nc}=p_{i}+\frac{1}{4\hbar}\epsilon_{ijk}\eta_{k}x_{j}:\begin{cases}
\begin{array}{ccc}
p_{x}^{nc} & = & p_{x}+\frac{1}{4\hbar}\eta y\\
p_{y}^{nc} & = & p_{y}-\frac{1}{4\hbar}\eta x\\
p_{z}^{nc} & = & p_{z}
\end{array}\end{cases}.\label{eq:5-1}
\end{equation}

$\phantom{}$

In noncommutative quantum mechanics, it is quite possible that we
replace the usual product with the Moyal-Weyl ($\star$) product,
then the quantum mechanical system will become simply the noncommutative
quantum mechanical system. Let $\mathcal{H}\left(x,p\right)$ be the
Hamiltonian operator of the usual quantum system, then the corresponding
Schrödinger equation on noncommutative quantum mechanics is typically
written as
\begin{equation}
\mathcal{H}\left(x,p\right)\star\psi\left(x,p\right)=E\psi\left(x,p\right).\label{eq:6}
\end{equation}

The definition of Moyal-Weyl product between two arbitrary functions
$f(x,p)$ and $g(x,p)$ in phase-space is given by \cite{key-37}
\begin{equation}
\begin{array}{c}
(f\star g)(x,p)=\exp[\frac{i}{2}\Theta_{ab}\partial_{x_{a}}\partial_{x_{b}}+\frac{i}{2}\eta_{ab}\partial_{p_{a}}\partial_{p_{b}}]f\left(x_{a},p_{a}\right)g\left(x_{b},p_{b}\right)=f(x,p)g(x,p)\\
+\sum_{n=1}\frac{1}{n!}\left(\frac{i}{2}\right)^{n}\Theta^{a_{1}b_{1}}...\Theta^{a_{n}b_{n}}\partial_{a_{1}}^{x}...\partial_{a_{k}}^{x}f(x,p)\partial_{b_{1}}^{x}...\partial_{b_{k}}^{x}g(x,p)\\
+\sum_{n=1}\frac{1}{n!}\left(\frac{i}{2}\right)^{n}\eta^{a_{1}b_{1}}...\eta^{a_{n}b_{n}}\partial_{a_{1}}^{p}...\partial_{a_{k}}^{p}f(x,p)\partial_{b_{1}}^{p}...\partial_{b_{k}}^{p}g(x,p)
\end{array},\label{eq:7}
\end{equation}
with $f(x,p)$ and $g(x,p)$, assumed to be infinitely differentiable.
If we consider the case of noncommutative space the definition of
Moyal-Weyl product will be reduced to \cite{key-38}

\begin{equation}
(f\star g)(x)=\exp[\frac{i}{2}\Theta_{ab}\partial_{x_{a}}\partial_{x_{b}}]f\left(x_{a}\right)g\left(x_{b}\right)=f(x)g(x)+\sum_{n=1}\frac{1}{n!}\left(\frac{i}{2}\right)^{n}\Theta^{a_{1}b_{1}}...\Theta^{a_{n}b_{n}}\partial_{a_{1}}...\partial_{a_{k}}f(x)\partial_{b_{1}}...\partial_{b_{k}}g(x).\label{eq:8}
\end{equation}

Due to the nature of the $\star$product, the noncommutative field
theories for low-energy fields ({\small{}$E^{2}\apprle1/\Theta$})
at classical level are completely reduced to their commutative versions.
However, this is just the classical result and quantum corrections
always reveal the effects of $\Theta$ even at low-energies.

$\phantom{}$

On noncommutative phase-space the $\star$product can be replaced
by a Bopp's shift, i.e. the $\star$product can be changed into the
ordinary product by replacing $\mathcal{H}\left(x,p\right)$ with
$\mathcal{H}\left(x^{nc},p^{nc}\right)$. Thus the corresponding noncommutative
Schrödinger equation can be written as
\begin{equation}
\mathcal{H}\left(x,p\right)\star\psi\left(x,p\right)=\mathcal{H}\left(x_{i}-\frac{1}{2\hbar}\Theta_{ij}p_{j},\;p_{\mu}+\frac{1}{2\hbar}\eta_{\mu\nu}x_{\nu}\right)\psi=E\psi.\label{eq:9}
\end{equation}

Note that $\Theta$ and $\eta$ terms always can be treated as a perturbation
in quantum mechanics.

If $\Theta=\eta=0$, the noncommutative algebra reduces to the ordinary
commutative one.

\section{Pauli equation in noncommutative phase-space }

\subsection{Formulation of noncommutative Pauli equation}

The Pauli equation is the formulation of the Schrödinger equation
for spin-1/2 particles, which was formulated by W. Pauli in 1927.
It takes into account the interaction of the particle's spin with
an electromagnetic field. However, in other words it is the nonrelativistic
limit of the Dirac equation. Besides, the Pauli equation could be
extracted from other relativistic higher spin equations such as the
DKP equation once considering the particle interacting with an electromagnetic
field \cite{key-37}. The nonrelativistic Schrödinger equation that
describes an electron in interaction with an electromagnetic potential
$\left(A_{0},\overrightarrow{A}\right)$ ($\hat{\overrightarrow{p}}$
is replaced with $\hat{\overrightarrow{\pi}}=\hat{\overrightarrow{p}}-\frac{e}{c}\overrightarrow{A}$
and $\hat{E}$ with $\hat{\epsilon}=i\hbar\frac{\partial}{\partial t}-e\phi$
) is\foreignlanguage{french}{
\begin{equation}
\frac{1}{2m}\left(\hat{\overrightarrow{p}}-\frac{e}{c}\overrightarrow{A}\left(r\right)\right)^{2}\psi\left(r,t\right)+e\phi\left(r\right)\psi\left(r,t\right)=i\hbar\frac{\partial}{\partial t}\psi\left(r,t\right),\label{eq:10}
\end{equation}
}where $\hat{\overrightarrow{p}}=i\hbar\overrightarrow{\nabla}$ is
the momentum operator, $m$, $e$ are the mass and charge of the electron,
and $c$ is the speed of light. $\psi\left(r,t\right)$ is Schrödinger\textquoteright s
scalar wave function. The appearance of real-valued electromagnetic
Coulomb and vector potentials, $\phi\left(\overrightarrow{r},t\right)$
and $\overrightarrow{A}\left(\overrightarrow{r},t\right)$, is a consequence
of using the gauge-invariant minimal coupling assumption to describe
the interaction with the external magnetic and electric fields defined
by 
\begin{equation}
\overrightarrow{E}=-\overrightarrow{\nabla}\phi-\frac{1}{c}\frac{\partial\overrightarrow{A}}{\partial t},\quad\overrightarrow{B}=\overrightarrow{\nabla}\times\overrightarrow{A}.\label{eq:11}
\end{equation}

However, the electron gains potential energy when the spin interacts
with the magnetic field, therefore the Pauli equation of an electron
with spin is given by \cite{key-1,key-8} 
\begin{equation}
\frac{1}{2m}\left(\overrightarrow{\sigma}.\hat{\overrightarrow{\pi}}\right)^{2}\psi\left(r,t\right)+e\phi\psi\left(r,t\right)=\frac{1}{2m}\left(\hat{\overrightarrow{p}}-\frac{e}{c}\overrightarrow{A}\right)^{2}\psi\left(r,t\right)+e\phi\psi\left(r,t\right)+\mu_{B}\overrightarrow{\sigma}.\overrightarrow{B}\psi\left(r,t\right)=i\hbar\frac{\partial}{\partial t}\psi\left(r,t\right),\label{eq:12}
\end{equation}
where $\psi\left(r,t\right)=\left(\begin{array}{cc}
\psi_{1} & \psi_{2}\end{array}\right)^{T}$ is the spinor wave function, which replaces the scalar wave function.
With $\mu_{B}=\frac{\left|e\right|\hbar}{2mc}=9.27\times10^{-24}JT^{-1}$
is Bohr's magneton, $\overrightarrow{B}$ is the applied magnetic
field vector, also $\mu_{B}\overrightarrow{\sigma}$ represents the
magnetic moment. $\overrightarrow{\sigma}$'s being the three Pauli
matrices ($\text{Tr}\overrightarrow{\sigma}=0$), obey the following
algebra
\begin{equation}
\left[\sigma_{i},\sigma_{j}\right]=2i\epsilon_{ijk}\sigma_{k},\label{eq:13}
\end{equation}
\begin{equation}
\sigma_{i}\sigma_{j}=\delta_{ij}I+i\sum_{k}\epsilon_{ijk}\sigma_{k},\label{eq:15-1}
\end{equation}
\begin{equation}
\left(\overrightarrow{\sigma}.\hat{\overrightarrow{a}}\right)\left(\overrightarrow{\sigma}.\hat{\overrightarrow{b}}\right)=\hat{\overrightarrow{a}}.\hat{\overrightarrow{b}}+i\overrightarrow{\sigma}.\left(\hat{\overrightarrow{a}}\times\hat{\overrightarrow{b}}\right),\label{eq:14-1}
\end{equation}
$\hat{\overrightarrow{a}}$, $\hat{\overrightarrow{b}}$ are any two
vector operators that commute with $\overrightarrow{\sigma}$. It
must be emphasized that the third term of equation (\ref{eq:12})
is the Zeeman term, which is generated automatically by using feature
(\ref{eq:14-1}) with a correct g-factor of $g=2$ as reduced in the
Bohr's magneton rather than being introduced by hand as a phenomenological
term, as is usually done.

The Pauli equation in noncommutative phase-space is
\begin{equation}
\mathcal{H}\left(x^{nc},p^{nc}\right)\psi\left(x^{nc},t\right)=\mathcal{H}\left(x,p^{nc}\right)\star\psi\left(x,t\right)=e^{\frac{i}{2}\Theta_{ab}\partial_{x_{a}}\partial_{x_{b}}}\mathcal{H}\left(x_{a},p^{nc}\right)\psi\left(x_{b},t\right)=i\hbar\frac{\partial}{\partial t}\psi\left(x,t\right).\label{eq:14}
\end{equation}

Here we achieved the noncommutativity in space using Moyal $\star$product
then the noncommutativity in phase through Bopp-shift. Using equation
(\ref{eq:8}), we have
\begin{equation}
\mathcal{H}\left(x^{nc},p^{nc}\right)\psi\left(x^{nc},t\right)=\left\{ \mathcal{H}\left(x,p^{nc}\right)+\frac{i}{2}\Theta^{ab}\partial_{a}\mathcal{H}\left(x,p^{nc}\right)\partial_{b}+\sum_{n=2}\frac{1}{n!}\left(\frac{i}{2}\right)^{n}\Theta^{a_{1}b_{1}}...\Theta^{a_{n}b_{n}}\partial_{a_{1}}...\partial_{a_{k}}\mathcal{H}\left(x,p^{nc}\right)\partial_{b_{1}}...\partial_{b_{k}}\right\} \psi.\label{eq:15}
\end{equation}

In case of a constant real magnetic field $\overrightarrow{B}=\left(0,0,B\right)=B\overrightarrow{e}_{3}$
oriented along the axis (Oz), which is often referred to as the Landau
system. We have the following symmetric gauge
\begin{equation}
\overrightarrow{A}=\frac{\overrightarrow{B}\times\overrightarrow{r}}{2}=\frac{B}{2}\left(-y,x,0\right),\;\text{with }\:A_{0}\left(x\right)=e\phi=0.\label{eq:16}
\end{equation}

Therefore, the derivations in the equation (\ref{eq:15}) approximately
shut down in the first-order of $\Theta$, then the noncommutative
Pauli equation in the presence of a uniform magnetic field can be
written as follows 
\begin{equation}
\mathcal{H}\left(x,p^{nc}\right)\star\psi\left(x\right)=\left\{ \frac{1}{2m}\left(\overrightarrow{p}^{nc}-\frac{e}{c}\overrightarrow{A}\left(x\right)\right)^{2}+\mu_{B}\overrightarrow{\sigma}.\overrightarrow{B}+\frac{ie}{4mc}\Theta^{ab}\partial_{a}\left(\frac{e}{c}\overrightarrow{A}^{2}-2\overrightarrow{p}^{nc}.\overrightarrow{A}\right)\partial_{b}\right\} \psi\left(x\right)+0(\Theta^{2}),\label{eq:17}
\end{equation}
with $\left[\overrightarrow{p}^{nc},\overrightarrow{A}\right]=0$.
We now make use of the Bopp-shift transformation (\ref{eq:4}), in
the momentum operator to obtain
\begin{equation}
\begin{array}{c}
\mathcal{H}\left(x^{nc},p^{nc}\right)\psi\left(x^{nc},t\right)=\left\{ \frac{1}{2m}\left(p_{i}+\frac{1}{2\hbar}\eta_{ij}x_{j}-\frac{e}{c}A_{i}\right)^{2}+\mu_{B}\overrightarrow{\sigma}.\overrightarrow{B}\right.\\
\left.-\frac{ie}{4mc}\Theta^{ab}\partial_{a}\left(2\left(p_{i}+\frac{1}{2\hbar}\eta_{ij}x_{j}\right)A_{i}-\frac{e}{c}\overrightarrow{A}^{2}\right)\partial_{b}\right\} \psi(x,t)=i\hbar\frac{\partial}{\partial t}\psi\left(x,t\right),
\end{array}\label{eq:18}
\end{equation}
we rewrite the above equation in a more compact form
\begin{equation}
\begin{array}{c}
\mathcal{H}\left(x,p^{nc}\right)\star\psi\left(x,t\right)=\left\{ \frac{1}{2m}\left(\overrightarrow{p}-\frac{e}{c}\overrightarrow{A}\right)^{2}-\frac{1}{2m}\left(\overrightarrow{x}\times\overrightarrow{p}\right).\overrightarrow{\eta}-\frac{1}{2m}\frac{e}{c\hbar}\left(\overrightarrow{x}\times\overrightarrow{A}\left(x\right)\right).\overrightarrow{\eta}\right.\\
\left.+\frac{1}{8m\hbar^{2}}\eta_{ij}\eta_{\alpha\beta}x_{j}x_{\beta}+\mu_{B}\overrightarrow{\sigma}.\overrightarrow{B}+\frac{e}{4\hbar mc}\left(\overrightarrow{\nabla}\left(2\overrightarrow{p}.\overrightarrow{A}-\frac{1}{2\hbar}\left(\overrightarrow{x}\times\overrightarrow{A}\left(x\right)\right).\overrightarrow{\eta}-\frac{e}{c}\overrightarrow{A}^{2}\right)\times\overrightarrow{p}\right).\overrightarrow{\Theta}\right\} \psi\left(x,t\right).
\end{array}\label{eq:19}
\end{equation}

We restrict ourselves only to the first-order of the parameter $\eta$.
The only reason behind this consideration is the balance with the
noncommutativity in the space considered in the case of constant magnetic
field. Thus we have now 
\begin{equation}
\begin{array}{c}
\mathcal{H}\left(x,p^{nc}\right)\star\psi\left(x,t\right)=\left\{ \frac{1}{2m}\left(\overrightarrow{p}-\frac{e}{c}\overrightarrow{A}\right)^{2}-\frac{1}{2m}\overrightarrow{L}.\overrightarrow{\eta}-\frac{e}{2mc\hbar}\left(\overrightarrow{x}\times\overrightarrow{A}\left(x\right)\right).\overrightarrow{\eta}+\mu_{B}\overrightarrow{\sigma}.\overrightarrow{B}\right.\\
\left.+\frac{e}{4mc\hbar}\left(\overrightarrow{\nabla}\left(2\overrightarrow{p}.\overrightarrow{A}\left(x\right)-\frac{1}{2\hbar}\left(\overrightarrow{x}\times\overrightarrow{A}\left(x\right)\right).\overrightarrow{\eta}-\frac{e}{c}\overrightarrow{A}^{2}\right)\times\overrightarrow{p}\right).\overrightarrow{\Theta}\right\} \psi\left(x,t\right))=i\hbar\frac{\partial}{\partial t}\psi\left(x,t\right).
\end{array}\label{eq:20}
\end{equation}

The existence of a Pauli equation for all orders of $\Theta$ parameter
is explicitly relative to the magnetic field.

In the case of a non-constant magnetic field, we introduce a function
depending on $x$ in the Landau gauge as $A_{2}=xBf(x)$ which gives
us a non-constant magnetic field. The magnetic field can be calculated
easily using the second equation of equation (\ref{eq:11}) as follows
\cite{key-33}
\begin{equation}
\overrightarrow{B}(x)=Bf(x)\overrightarrow{e}_{3}.\label{eq:21}
\end{equation}

If we specify $f(x)$ we obtain different classes of the non-constant
magnetic field. If take $f(x)=1$ in this case we get a constant magnetic
field. 

Having the equation (\ref{eq:20}) on hand, we calculate the probability
density and the current density.

\subsection{Deformed continuity equation}

In the following we calculate the current density, which results from
the Pauli equation (\ref{eq:20} that describing a system of two coupled
differential equations for $\psi_{1}$ and $\psi_{2}$.

By putting 
\begin{equation}
\begin{array}{cc}
\mathcal{Q}_{\eta}=\mathcal{Q}_{\eta}^{\ast}=\left(\overrightarrow{x}\times\overrightarrow{A}\left(x\right)\right).\overrightarrow{\eta},\; & \mathcal{Q}_{\Theta}=\left(\overrightarrow{\nabla}\left(2\overrightarrow{p}.\overrightarrow{A}\left(x\right)-\frac{1}{2\hbar}\mathcal{Q}_{\eta}-\frac{e}{c}\overrightarrow{A}^{2}\left(x\right)\right)\times\overrightarrow{p}\right).\overrightarrow{\Theta}\end{array}=\left(\overrightarrow{\nabla}\mathcal{V}\left(x\right)\times\overrightarrow{p}\right).\overrightarrow{\Theta},\label{eq:22}
\end{equation}
the noncommutative Pauli equation in the presence of a uniform magnetic
field simply reads
\begin{equation}
\left\{ \frac{1}{2m}\left(-\hbar^{2}\overrightarrow{\nabla}^{2}+\frac{ie\hbar}{c}\left(\overrightarrow{\nabla}.\overrightarrow{A}+\overrightarrow{A}.\overrightarrow{\nabla}\right)+\frac{e^{2}}{c^{2}}\overrightarrow{A}^{2}\right)-\frac{\overrightarrow{L}.\overrightarrow{\eta}}{2m}-\frac{e\mathcal{Q}_{\eta}}{2mc\hbar}+\mu_{B}\overrightarrow{\sigma}.\overrightarrow{B}+\frac{e\mathcal{Q}_{\Theta}}{4mc\hbar}\right\} \psi=i\hbar\frac{\partial}{\partial t}\psi.\label{eq:23}
\end{equation}

Knowing that $\overrightarrow{\sigma}$, $\overrightarrow{L}$ are
Hermitian and the magnetic field is real, and $\mathcal{Q}_{\Theta}^{\ast}$
is the adjoint of $\mathcal{Q}_{\Theta}$. 

The adjoint equation of equation (\ref{eq:23}) reads
\begin{equation}
\frac{1}{2m}\left\{ -\hbar^{2}\overrightarrow{\nabla}^{2}\psi^{\dagger}-\frac{ie\hbar}{c}\left(\overrightarrow{\nabla}.\overrightarrow{A}+\overrightarrow{A}.\overrightarrow{\nabla}\right)\psi^{\dagger}+\frac{e^{2}}{c^{2}}\overrightarrow{A}^{2}\psi^{\dagger}\right\} -\frac{\overrightarrow{L}.\overrightarrow{\eta}}{2m}\psi^{\dagger}-\frac{e\mathcal{Q}_{\eta}}{2mc\hbar}\psi^{\dagger}+\mu_{B}\overrightarrow{\sigma}.\overrightarrow{B}\psi^{\dagger}+\frac{e}{4mc\hbar}\psi^{\dagger}\mathcal{Q}_{\Theta}^{\ast}=-i\hbar\frac{\partial\psi^{\dagger}}{\partial t}.\label{eq:24}
\end{equation}

Here $\ast$, $\text{\dag}$ stand for the complex conjugation of
the potentials, operators and for the wave-functions successively. 

To find the continuity equation, we multiply equation (\ref{eq:23})
from left by $\psi^{\dagger}$ and equation (\ref{eq:24}) from the
right by $\psi$, then making the subtraction of these equations,
yields

\begin{equation}
\begin{array}{c}
\frac{-\hbar^{2}}{2m}\left\{ \psi^{\dagger}\overrightarrow{\nabla}^{2}\psi-\left(\overrightarrow{\nabla}^{2}\psi^{\dagger}\right)\psi\right\} +\frac{ie\hbar}{2mc}\left\{ \psi^{\dagger}\left(\overrightarrow{\nabla}.\overrightarrow{A}+\overrightarrow{A}.\overrightarrow{\nabla}\right)\psi+\left[\left(\overrightarrow{\nabla}.\overrightarrow{A}+\overrightarrow{A}.\overrightarrow{\nabla}\right)\psi^{\dagger}\right]\psi\right\} \\
+\frac{e}{4mc\hbar}\left(\psi^{\dagger}\mathcal{Q}_{\Theta}\psi-\psi^{\dagger}\mathcal{Q}_{\Theta}^{\ast}\psi\right)=i\hbar\left(\psi^{\dagger}\frac{\partial}{\partial t}\psi+\psi\frac{\partial}{\partial t}\psi^{\dagger}\right),
\end{array}\label{eq:25}
\end{equation}
after some minor simplefications, we have 

\begin{equation}
\frac{-\hbar}{2m}div\left\{ \psi^{\dagger}\overrightarrow{\nabla}\psi-\psi\overrightarrow{\nabla}\psi^{\dagger}\right\} +\frac{ie}{mc}div\left\{ \overrightarrow{A}\psi^{\dagger}\psi\right\} +\frac{e}{4mc\hbar^{2}}\left(\psi^{\dagger}\mathcal{Q}_{\Theta}\psi-\psi^{\dagger}\mathcal{Q}_{\Theta}^{\ast}\psi\right)=i\frac{\partial}{\partial t}\psi^{\dagger}\psi.\label{eq:26}
\end{equation}

This will be recognized as the deformed continuity equation. The obtained
equation (\ref{eq:26}) contains new quantity, which is the deformation
due to the effect of the phase-space noncommutativity on the Pauli
equation.

The third term on the left-hand side, which is the deformation quantity,
can be simplified as follows
\begin{equation}
\frac{ie}{4mc\hbar^{2}}\left(\psi^{\dagger}\mathcal{Q}_{\Theta}\psi-\psi^{\dagger}\mathcal{Q}_{\Theta}^{\ast}\psi\right)=\frac{ie}{4mc\hbar^{2}}\left(\psi^{\dagger}\left(\mathcal{V}\left(x\right)\star\psi\right)-\left(\psi^{\dagger}\star\mathcal{V}\left(x\right)\right)\psi\right),\label{eq:27}
\end{equation}
using the propriety $\left(\overrightarrow{a}\times\overrightarrow{b}\right).\overrightarrow{c}=\overrightarrow{a}.\left(\overrightarrow{b}\times\overrightarrow{c}\right)=\overrightarrow{b}.\left(\overrightarrow{c}\times\overrightarrow{a}\right)$,
also we must pay attention to the order, $\psi^{\dagger}$ is the
first and $\psi$ the second factor, we have
\begin{equation}
\frac{ie}{4mc\hbar^{2}}\left(\psi^{\dagger}\mathcal{Q}_{\Theta}\psi-\psi^{\dagger}\mathcal{Q}_{\Theta}^{\ast}\psi\right)=\frac{e}{8mc\hbar^{2}}div\mathcal{V}\left(x\right)\left(\overrightarrow{\Theta}\times\overrightarrow{\nabla}\left(\psi^{\dagger}\psi\right)\right)=div\overrightarrow{\xi}^{nc}.\label{eq:28}
\end{equation}

Using the following identity also gives the same equation above \cite{key-6}
\begin{equation}
\upsilon^{\dagger}\left(\overrightarrow{\pi}\tau\right)-\left(\overrightarrow{\pi}\upsilon\right)^{\dagger}\tau=-i\hbar\overrightarrow{\nabla}\left(\upsilon^{\dagger}\tau\right),\label{eq:29}
\end{equation}
where $\upsilon$, $\tau$ are arbitrary two-component spinor. Noting
that $\overrightarrow{A}$ does not appear on the right-hand side
of the identity; and that this identity is related to the fact that
$\overrightarrow{\pi}$ is Hermitian. 

It is evident that the noncommutativity affects the current density,
and the deformation quantity may apear as a correction to it. The
deformed current density satisfies the current conservation, which
means, we have a conservation of the continuity equation in the noncommutative
phase-space. Equation (\ref{eq:26}) may be contracted as
\begin{equation}
\frac{\partial\rho}{\partial t}+\overrightarrow{\nabla}.\overrightarrow{j}^{nc}=0,\label{eq:30}
\end{equation}
where
\begin{equation}
\rho=\psi^{\dagger}\psi=\left|\psi\right|^{2},\label{eq:31}
\end{equation}
is the probability density and 
\begin{equation}
\overrightarrow{j}^{nc}=\overrightarrow{j}+\overrightarrow{\xi}^{nc}=\frac{-i\hbar}{2m}\left\{ \psi^{\dagger}\overrightarrow{\nabla}\psi-\psi\overrightarrow{\nabla}\psi^{\dagger}\right\} -\frac{e}{mc}\left\{ \overrightarrow{A}\psi^{\dagger}\psi\right\} +\overrightarrow{\xi}^{nc},\label{eq:32}
\end{equation}
is the deformed current density of the electrons. The deformation
quantity is
\begin{equation}
\overrightarrow{\xi}^{nc}=\frac{e}{8mc\hbar^{2}}\mathcal{V}\left(x\right)\left(\overrightarrow{\Theta}\times\overrightarrow{\nabla}\left(\psi^{\dagger}\psi\right)\right)=\frac{e}{8mc\hbar^{2}}\left(2\overrightarrow{p}.\overrightarrow{A}-\frac{1}{2\hbar}\mathcal{Q}_{\eta}-\frac{e}{c}\overrightarrow{A}^{2}\right)\left(\overrightarrow{\Theta}\times\overrightarrow{\nabla}\left(\psi^{\dagger}\psi\right)\right).\label{eq:33}
\end{equation}

The existence of a deformed continuity equation for all orders of
$\Theta$ parameter also proportional to the magnetic field. Actually,
one can explicitly calculate the conserved current to all orders of
$\Theta$. In the case of a non-constant magnetic field, and using
equation (\ref{eq:21}), we have
\begin{equation}
\frac{\partial\rho}{\partial t}+\overrightarrow{\nabla}.\overrightarrow{j}+\frac{ie}{4mc\hbar^{2}}\left\{ \psi^{\dagger}\left(\mathcal{V}\left(x\right)\star\psi\right)-\left(\psi^{\dagger}\star\mathcal{V}\left(x\right)\right)\psi\right\} =0,\label{eq:34}
\end{equation}
we calculate the $n^{th}$ order term in the general deformed continuity
equation (\ref{eq:34}) as follows
\begin{equation}
\begin{array}{c}
\left.\psi^{\dagger}\left(\mathcal{V}\left(x\right)\star\psi\right)-\left(\psi^{\dagger}\star\mathcal{V}\left(x\right)\right)\psi\right|_{n^{th}}=\frac{1}{n!}\left(\frac{i}{2}\right)^{n}\Theta^{a_{1}b_{1}}...\Theta^{a_{n}b_{n}}\\
\times\left(\psi^{\dagger}\partial_{a_{1}}...\partial_{a_{k}}\mathcal{V}\left(x\right)\partial_{b_{1}}...\partial_{b_{k}}\psi-\partial_{a_{1}}...\partial_{a_{k}}\psi^{\dagger}\left(\partial_{b_{1}}...\partial_{b_{k}}\mathcal{V}\left(x\right)\right)\psi+(-1)^{n}cc.\right).
\end{array}\label{eq:35}
\end{equation}

We note the absence of the magnetization current term in equation
(\ref{eq:32}), as in commutative case when this was asserted by authors
\cite{key-1,key-4,key-8,key-13,key-16,key-27}, where at first they
attempted to cover this deficiency by explaining how to derive this
additional term from the non-relativistic limit of the relativistic
Dirac probability current density. Then, Nowakowski and others \cite{key-6}
provided a superb explanation of how to extract this term through
the non-relativistic Pauli equation itself. 

Knowing that, in commutative background the magnetization current
$\overrightarrow{j}_{M}$ from the probability current of Pauli equation
is proportional to $\overrightarrow{\nabla}\times\left(\psi^{\dagger}\overrightarrow{\sigma}\psi\right)$.
However, the existence of such an additional term is important and
it should be discussed when talking about the probability current
of spin-1/2 particles. In following, we try to derive the current
magnetization in noncommutative background without changing the continuity
equation, and seek if such additional term is affected by noncommutativity
or not.

\subsection{Derivation of the magnetization current}

At first it must be clarified that the authors Nowakowski and others
(2011) in \cite{key-4,key-6} derived the non-relativistic current
density for a spin-1/2 particle using \textbf{minimally coupled Pauli
equation}. In contrast, Wilkes, J. M (2020) in \cite{key-39} derived
the non-relativistic current density for a free spin-1/2 particle
using directly\textbf{ free Pauli equatio}n. However, we show here
that the current density can be derived from the minimally coupled
Pauli equation in noncommutative phase-space. 

Starting with the \textbf{noncommutative minimally coupled Pauli}
equation written in the form 
\begin{equation}
\mathcal{H}_{Pauli}^{nc}\psi=\frac{1}{2m}\left(\overrightarrow{\sigma}.\hat{\overrightarrow{\pi}}^{nc}\right)^{2}\psi=i\hbar\frac{\partial}{\partial t}\psi,\label{eq:38-1}
\end{equation}
we multiply the above equation from left by $\psi^{\dagger}$ and
the adjoint equation of equation (\ref{eq:38-1}) from the right by
$\psi$, the subtraction of these equations yields the following continuity
equation

\begin{equation}
2m\left\{ \left(\left(\overrightarrow{\sigma}.\hat{\overrightarrow{\pi}}^{nc}\right)^{2}\psi\right)^{\dagger}\psi-\psi^{\dagger}\left(\overrightarrow{\sigma}.\hat{\overrightarrow{\pi}}^{nc}\right)^{2}\psi\right\} =i\hbar\left(\psi^{\dagger}\frac{\partial\psi}{\partial t}+\psi\frac{\partial\psi^{\dagger}}{\partial t}\right),\label{eq:40-1}
\end{equation}
noting that the noncommutativity of $\pi^{nc}$ has led us to express
the two terms as follows
\begin{equation}
\frac{i}{2m\hbar}\sum_{i,j}\left\{ \left(\hat{\pi_{i}}^{nc}\hat{\pi_{j}}^{nc}\psi\right)^{\dagger}\sigma_{j}\sigma_{i}\psi-\psi^{\dagger}\sigma_{i}\sigma_{j}\left(\hat{\pi_{i}}^{nc}\hat{\pi_{j}}^{nc}\psi\right)\right\} =\frac{\partial\rho}{\partial t}.\label{eq:41-1}
\end{equation}

While with only $p_{i}$, we would have no reason for preferring $p_{i}p_{j}\psi$
over $p_{j}p_{i}\psi$. 

It is easy to verify that the identity (\ref{eq:29}) remains valid
for $\overrightarrow{\pi}^{nc}$ because the fact that $\overrightarrow{\pi}^{nc}$
is Hermitian. Therefore, through identity (\ref{eq:29}), we have
\begin{equation}
\frac{-1}{2m}\sum_{i,j}\nabla_{i}\left\{ \left(\hat{\pi_{j}}^{nc}\psi\right)^{\dagger}\sigma_{j}\sigma_{i}\psi+\psi^{\dagger}\sigma_{i}\sigma_{j}\left(\hat{\pi_{j}}^{nc}\psi\right)\right\} +\frac{i}{2m\hbar}\sum_{i,j}\left\{ \left(\hat{\pi_{j}}^{nc}\psi\right)^{\dagger}\sigma_{j}\sigma_{i}\left(\hat{\pi_{i}}^{nc}\psi\right)-\left(\hat{\pi_{i}}^{nc}\psi\right)^{\dagger}\sigma_{i}\sigma_{j}\left(\hat{\pi_{j}}^{nc}\psi\right)\right\} =\frac{\partial\rho}{\partial t},\label{eq:42-1}
\end{equation}
then
\begin{equation}
\frac{-1}{2m}\sum_{i,j}\nabla_{i}\left\{ \left(\hat{\pi_{j}}^{nc}\psi\right)^{\dagger}\sigma_{j}\sigma_{i}\psi+\psi^{\dagger}\sigma_{i}\sigma_{j}\left(\hat{\pi_{j}}^{nc}\psi\right)\right\} =\frac{\partial\rho}{\partial t}.\label{eq:43-1}
\end{equation}

Knowing that the $2{}^{nd}$ sum in equation (\ref{eq:42-1}) gives
zero by swapping $i$ and $j$ for one of the sums, then the probability
current vector from the above continuity equation is
\begin{equation}
j_{i}=\frac{1}{2m}\sum_{j}\left\{ \left(\hat{\pi_{j}}^{nc}\psi\right)^{\dagger}\sigma_{j}\sigma_{i}\psi+\psi^{\dagger}\sigma_{i}\sigma_{j}\left(\hat{\pi_{j}}^{nc}\psi\right)\right\} .\label{eq:44-1}
\end{equation}

Using the property (\ref{eq:15-1}), equation (\ref{eq:44-1}) becomes
\begin{equation}
j_{i}=\frac{1}{2m}\sum_{j}\left\{ \left(\hat{\pi_{j}}^{nc}\psi\right)^{\dagger}\psi+\psi^{\dagger}\left(\hat{\pi_{j}}^{nc}\psi\right)+i\sum_{k}\left[\epsilon_{jik}\left(\hat{\pi_{j}}^{nc}\psi\right)^{\dagger}\sigma_{k}\psi+\epsilon_{ijk}\psi^{\dagger}\sigma_{k}\left(\hat{\pi_{j}}^{nc}\psi\right)\right]\right\} ,\label{eq:45-1}
\end{equation}
with $\epsilon_{jik}=-\epsilon_{ijk}$, and using one more time identity
(\ref{eq:29}), we find (this is similar to investigation by \cite{key-6}
in the case of commutative phase-space) 
\begin{equation}
j_{i}=\frac{1}{2m}\left[\left(\hat{p_{j}}^{nc}\psi\right)^{\dagger}\psi-\frac{e}{c}\left(A_{j}^{nc}\psi\right)^{\dagger}\psi+\psi^{\dagger}\hat{p_{j}}^{nc}\psi-\frac{e}{c}\psi^{\dagger}A_{j}^{nc}\psi\right]+\frac{\hbar}{2m}\sum_{j,k}\epsilon_{ijk}\nabla_{j}\left(\psi^{\dagger}\sigma_{k}\psi\right).\label{eq:46-1}
\end{equation}

In the right-hand side of the above equation, the first term will
be interpreted as the noncommutative current vector $\overrightarrow{j}^{nc}$
given by equation (\ref{eq:33}), and the second term is the requested
additional term, namely current magnetization $\overrightarrow{j}_{M}$,
where 
\begin{equation}
\underset{M}{j}_{i}=\frac{\hbar}{2m}\left(\overrightarrow{\nabla}\times\left(\psi^{\dagger}\overrightarrow{\sigma}\psi\right)\right)_{i}.\label{eq:47-1}
\end{equation}

Besides, $\overrightarrow{j}_{M}$ can also be shown to be a part
of the conserved Noether current \cite{key-40}, resulting from the
invariance of the Pauli Lagrangian under the global phase transformation
U(1). 

What can be concluded here is that the magnetization current is not
affected by the noncommutativity, perhaps because the spin operator
could not be affected by noncommutativity. This is in contrast to
what was previously found around the current density, which showed
a great influence of noncommutativity.

\section{Noncommutative Semi-classical Partition Function}

In this part of our work, we investigate the magnetization and the
magnetic susceptibility quantities of our Pauli system using the partition
function in noncommutative phase-space. We concentrate, at first,
on the calculation of the semi-classical partition function. Our studied
system is semi-classical, so our system is not completely classical
but contains a quantum interaction concerning the spin, therefore,
the noncommutative partition function is separable into two independent
parts as follows 
\begin{equation}
\mathcal{Z}^{nc}=Z_{clas}^{nc}Z_{ncl},\label{eq:48-1}
\end{equation}
where $Z_{ncl}$ is the non-classical part of the partition function.
To study our noncommutative classical partition function, we assume
that the passage between noncommutative classical mechanics and noncommutative
quantum mechanics can be realized through the following generalized
Dirac quantization condition \cite{key-42,key-43,key-44} 
\begin{equation}
\left\{ f,g\right\} =\frac{1}{i\hbar}\left[F,G\right],\label{eq:48-2}
\end{equation}
where $F$, $G$ stand for the operators associated with classical
observables $f$, $g$ and $\left\{ ,\right\} $ stands for Poisson
bracket. Using the condition above, we obtain from Eq.(\ref{eq:5})
\begin{equation}
\begin{array}{cccc}
\left\{ x_{j}^{nc},x_{k}^{nc}\right\} =\frac{1}{2}\epsilon_{jkl}\Theta_{l},\: & \left\{ p_{j}^{nc},p_{k}^{nc}\right\} =\frac{1}{2}\epsilon_{jkl}\eta_{l},\: & \left\{ x_{j}^{nc},p_{k}^{nc}\right\} =\delta_{jk}+\frac{1}{4}\Theta_{jl}\eta_{kl} & ,\:(j,k,l=1,2,3)\end{array}.\label{eq:48-3}
\end{equation}

Now based on the proposal that noncommutative observables $F^{nc}$
corresponding to the commutative one $F(x,p)$ can be defined by \cite{key-41,key-45}
\begin{equation}
F^{nc}=F(x^{nc},p^{nc}),\label{eq:48-4}
\end{equation}
 and for non-interacting particles, the classical partition function
in the canonical ensemble in noncommutative phase-space is given by
the following formula \cite{key-42,key-43} 
\begin{equation}
Z_{clas}^{nc}=\frac{1}{N!\left(2\pi\tilde{\hbar}\right)^{3N}}\int e^{-\beta\mathcal{H}_{clas}^{nc}\left(x,p\right)}d^{3N}x^{nc}d^{3N}p^{nc},\label{eq:36}
\end{equation}
which is written for $N$ particles. $\frac{1}{N!}$ is the Gibbs
correction factor, considered due to accounting for indistinguishability,
which means that there are $N!$ ways of arranging $N$ particles
at $N$ sites. $\tilde{\hbar}\sim\triangle x^{nc}\triangle p^{nc}$,
with $\frac{1}{\tilde{\hbar}^{3}}$ is a factor that makes the volume
of the noncommutative phase-space dimensionless. 

$\beta$ defined as $\frac{1}{K_{B}T}$, $K_{B}$ is the Boltzmann
constant, where $K_{B}=1.38\times10^{-23}JK^{-1}$. The Helmholtz
free energy is
\begin{equation}
F=-\frac{1}{\beta}\text{ln}Z,\label{eq:37}
\end{equation}
we may derive the magnetization as follows
\begin{equation}
\left\langle M\right\rangle =-\frac{\partial F}{\partial B}.\label{eq:38}
\end{equation}

For a single particle, the noncommutative classical partition function
is then
\begin{equation}
Z_{clas,1}^{nc}=\frac{1}{\tilde{h}^{3}}\int e^{-\beta\mathcal{H}_{Clas}^{nc}\left(x,p\right)}d^{3}x^{nc}d^{3}p^{nc},\label{eq:39}
\end{equation}
where $d^{3}$ is a shorthand notation serving as a reminder that
the $x$ and $p$ are vectors in three-dimensional phase-space. The
relation between equation (\ref{eq:36}) and (\ref{eq:39}) is given
by the following formula
\begin{equation}
Z_{clas}^{nc}=\frac{\left(Z_{clas,1}^{nc}\right)^{N}}{N!}.\label{eq:40}
\end{equation}

Knowing that using equation (\ref{eq:5-1}), we have 
\begin{equation}
d^{3}x^{nc}d^{3}p^{nc}=\left(1-\frac{\Theta\eta}{8\hbar^{2}}\right)d^{3}xd^{3}p,\label{eq:42}
\end{equation}
besides, using uncertainty principle and according to the third equation
of equation (\ref{eq:4}), we deduce 
\begin{equation}
\tilde{h}^{3}=h^{3}\left(1+\frac{3\Theta\eta}{4\hbar^{2}}\right)+\mathcal{O}\left(\Theta^{2}\eta^{2}\right).\label{eq:43}
\end{equation}
Unlike other works such as \cite{key-42}, when the researchers used
a different formula for Planck's constant $\tilde{h}_{11}=\tilde{h}_{22}\neq\tilde{h}_{33},$
which led to a different formula of $\tilde{h}^{3}$.

For an electron with spin in interaction with an electromagnetic potential,
once the magnetic field $\overrightarrow{B}$ be in the z-direction,
and by equation (\ref{eq:16}), bear in mind that $\left[\overrightarrow{p}^{nc},\overrightarrow{A}^{nc}\right]=0$,
then for the sake of simplicity, the noncommutative Pauli Hamiltonian
from equation (\ref{eq:20}) takes the form 
\begin{equation}
\mathcal{H}_{Pauli}\left(x^{nc},p^{nc}\right)=\frac{1}{2m}\left\{ \left(\overrightarrow{p}^{nc}\right)^{2}-2\frac{e}{c}\overrightarrow{p}^{nc}.\overrightarrow{A}^{nc}+\left(\frac{e}{c}\right)^{2}\left(\overrightarrow{A}^{nc}\right)^{2}\right\} +\mu_{B}\hat{\sigma}_{z}B.\label{eq:44}
\end{equation}

We split the noncommutative Pauli Hamiltonian as $\mathcal{H}_{Pauli}^{nc}=\mathcal{H}_{cla}^{nc}+\mathcal{H}_{ncl,\sigma}$,
with $\mathcal{H}_{ncl,\sigma}=\mu_{B}\hat{\sigma}_{z}B$ . 

It is easy to verify that
\begin{equation}
\left(\overrightarrow{p}^{nc}\right)^{2}=\left(p_{x}^{nc}\right)^{2}+\left(p_{y}^{nc}\right)^{2}+\left(p_{z}^{nc}\right)^{2}=p_{x}^{2}+p_{y}^{2}+p_{z}^{2}-\frac{\eta}{2\hbar}L_{z}+\frac{\eta^{2}}{16\hbar^{2}}\left(x^{2}+y^{2}\right),\label{eq:45}
\end{equation}
\begin{equation}
\overrightarrow{p}^{nc}.\overrightarrow{A}=p_{x}^{nc}A_{x}^{nc}+p_{y}^{nc}A_{y}^{nc}=\frac{B}{2}\left\{ -\frac{\Theta}{4\hbar}\left(p_{x}^{2}+p_{y}^{2}\right)-\frac{\eta}{4\hbar}\left(y^{2}+x^{2}\right)+\left(1+\frac{\Theta\eta}{16\hbar^{2}}\right)L_{z}\right\} ,\label{eq:46}
\end{equation}
\begin{equation}
\left(\overrightarrow{A}^{nc}\right)^{2}=\left(A_{x}^{nc}\right)^{2}+\left(A_{y}^{nc}\right)^{2}=\frac{B^{2}}{4}\left\{ x^{2}+y^{2}-\frac{\Theta}{2\hbar}L_{z}+\frac{\Theta^{2}}{16\hbar^{2}}\left(p_{x}^{2}+p_{y}^{2}\right)\right\} .\label{eq:47}
\end{equation}

Using the three equations above, our noncommutative classical Hamiltonian
becomes
\begin{equation}
\mathcal{H}_{cla}^{nc}=\frac{1}{2\tilde{m}}\left(p_{x}^{2}+p_{y}^{2}\right)+\frac{1}{2m}p_{z}^{2}-\tilde{\omega}L_{z}+\frac{1}{2}\tilde{m}\tilde{\omega}^{2}\left(x^{2}+y^{2}\right),\label{eq:48}
\end{equation}
where $L_{z}=p_{y}x-p_{x}y=\left(x_{i}\times p_{i}\right)_{z}$, and
\begin{equation}
\tilde{m}=\frac{m}{\left(1+\frac{eB\Theta}{8c\hbar}\right)^{2}},\text{ and }\tilde{\omega}=\frac{c\eta+2e\hbar B}{4c\hbar\tilde{m}\left(1+\frac{eB\Theta}{c8\hbar}\right)},\quad\frac{1}{2}\tilde{m}\tilde{\omega}^{2}=\frac{1}{2m}\left(\frac{\eta eB}{4c\hbar}+\frac{\eta^{2}}{16\hbar^{2}}+\frac{e^{2}B^{2}}{c^{2}4}\right).\label{eq:49}
\end{equation}

Now, following the definition given in equation (\ref{eq:39}) we
express the single particle noncommutative classical partition function
as {\large{}
\begin{equation}
Z_{clas,1}^{nc}=\frac{1}{\tilde{h}^{3}}\int e^{-\beta\left[\frac{1}{2\tilde{m}}\left(p_{x}^{2}+p_{y}^{2}\right)+\frac{1}{2m}p_{z}^{2}-\tilde{\omega}L_{z}+\frac{1}{2}\tilde{m}\tilde{\omega}^{2}\left(x^{2}+y^{2}\right)\right]}d^{3}x^{nc}d^{3}p^{nc}.\label{eq:50}
\end{equation}
}{\large \par}

It should be noted that once we want to factorize our Hamiltonian
into momentum and position terms.\textcolor{red}{{} }This is not always
possible when there are matrices (or operators) in the exponent. However,
within the classical limit, it is possible. Otherwise, to separate
the operators in the exponent, we use the Baker-Campbell-Hausdorff
(BCH) formula given by (first few terms) {\large{}
\begin{equation}
e^{\left[\hat{A}+\hat{B}\right]}=e^{\left[\hat{A}\right]}e^{\left[\hat{B}\right]}e^{\left[-\frac{1}{2}\left[\hat{A},\hat{B}\right]\right]}e^{\frac{1}{6}\left(2\left[\hat{A},\left[\hat{A},\hat{B}\right]\right]+\left[\hat{B},\left[\hat{A},\hat{B}\right]\right]\right)}...\label{eq:50-1}
\end{equation}
} 

We can now start to replace some of the operators in the exponent{\large{}
\begin{equation}
Z_{clas,1}^{nc}=\frac{1}{\tilde{h}^{3}}\int e^{-\beta\left[\frac{1}{2\tilde{m}}\left(p_{x}^{2}+p_{y}^{2}\right)+\frac{1}{2m}p_{z}^{2}\right]}e^{-\beta\left[\frac{1}{2}\tilde{m}\tilde{\omega}^{2}\left(x^{2}+y^{2}\right)\right]}e^{\beta\tilde{\omega}L_{z}}d^{3}p^{nc}d^{3}x^{nc}.\label{eq:62}
\end{equation}
}{\large \par}

We should expand exponentials containing $\tilde{\omega}$, and by
considering terms up to the second-order of $\tilde{\omega}$, we
obtain
\begin{equation}
Z_{clas,1}^{nc}=\frac{1}{\tilde{h}^{3}}\int e^{-\frac{\beta}{2}\left[\frac{p_{x}^{2}+p_{y}^{2}}{\tilde{m}}+\frac{p_{z}^{2}}{m}\right]}\left(1+\beta\tilde{\omega}L_{z}+\frac{1}{2}\beta^{2}\tilde{\omega}^{2}L_{z}^{2}\right)\left(1-\beta\tilde{\omega}^{2}\frac{\tilde{m}}{2}\left(x^{2}+y^{2}\right)\right)d^{3}p^{nc}d^{3}x^{nc},\label{eq:63}
\end{equation}
therefore, we have the appropriate expression for $Z_{clas,1}^{nc}${\large{}
\begin{equation}
\begin{array}{c}
Z_{clas,1}^{nc}=\frac{1-\frac{7\Theta\eta}{8\hbar^{2}}}{h^{3}}\int e^{-\frac{\beta}{2}\left[\frac{p_{x}^{2}+p_{y}^{2}}{\tilde{m}}+\frac{p_{z}^{2}}{m}\right]}d^{3}pd^{3}x+\frac{\left(1-\frac{7\Theta\eta}{8\hbar^{2}}\right)\beta\tilde{\omega}}{h^{3}}\int e^{-\frac{\beta}{2}\left[\frac{p_{x}^{2}+p_{y}^{2}}{\tilde{m}}+\frac{p_{z}^{2}}{m}\right]}L_{z}d^{3}pd^{3}x\\
+\frac{\left(1-\frac{7\Theta\eta}{8\hbar^{2}}\right)\beta^{2}\tilde{\omega}^{2}}{2h^{3}}\int e^{-\frac{\beta}{2}\left[\frac{p_{x}^{2}+p_{y}^{2}}{\tilde{m}}+\frac{p_{z}^{2}}{m}\right]}L_{z}^{2}d^{3}pd^{3}x-\frac{\left(1-\frac{7\Theta\eta}{8\hbar^{2}}\right)\beta\tilde{\omega}^{2}}{2h^{3}}\int e^{-\frac{\beta}{2}\left[\frac{p_{x}^{2}+p_{y}^{2}}{\tilde{m}}+\frac{p_{z}^{2}}{m}\right]}\left(x^{2}+y^{2}\right)d^{3}pd^{3}x.
\end{array}\label{eq:64}
\end{equation}
}{\large \par}

In the right-hand side of the above equation, it is easy to check
that the second integral goes to zero, the third and last integrals
cancel each other, and thus we obtain{\large{}
\begin{equation}
Z_{clas,1}^{nc}=\frac{1-\frac{7\Theta\eta}{8\hbar^{2}}}{h^{3}}\int e^{-\frac{\beta}{2}\left[\frac{p_{x}^{2}+p_{y}^{2}}{\tilde{m}}+\frac{p_{z}^{2}}{m}\right]}d^{3}pd^{3}x.\label{eq:65}
\end{equation}
}{\large \par}

Using the integral of Gaussian function $\int e^{-ax^{2}}dx=\sqrt{\frac{\pi}{a}}$,
we have {\large{}
\begin{equation}
Z_{clas,1}^{nc}=\frac{1-\frac{7\Theta\eta}{8\hbar^{2}}}{h^{3}}\int d^{3}x\int e^{-\frac{\beta}{2}\left[\frac{p_{x}^{2}+p_{y}^{2}}{\tilde{m}}+\frac{p_{z}^{2}}{m}\right]}d^{3}p=\frac{V}{\varLambda^{3}}\frac{1-\frac{7\Theta\eta}{8\hbar^{2}}}{\left(1+\frac{eB\Theta}{8c\hbar}\right)^{2}},\label{eq:66}
\end{equation}
}where $V$, $\varLambda=h\left(2m\pi K_{B}T\right)^{-\frac{1}{2}}$
are respectively the volume and the thermal de Broglie wavelength.
The non-classical partition function using $\mathcal{H}_{ncl,\sigma}$
is
\begin{equation}
Z_{ncl}=Z_{ncl,1}^{N}=\left(\sum_{\sigma_{z}=\pm1}e^{\beta\mu_{B}\hat{\sigma}_{z}B}\right)^{N}=2^{N}\text{cosh}{}^{N}\left(\beta\mu_{B}B\right).\label{eq:67}
\end{equation}

Finally, the Pauli partition function for a system of $N$ particles
in a three-dimensional noncommutative phase-space is 
\begin{equation}
\mathcal{Z}^{nc}=\frac{\left(2V\right)^{N}}{\varLambda^{3N}N!}\frac{\left(1-\frac{7\Theta\eta}{8\hbar^{2}}\right)^{N}\text{cosh}{}^{N}\left(\beta\mu_{B}B\right)}{\left(1+\frac{eB\Theta}{8c\hbar}\right)^{2N}}.\label{eq:68}
\end{equation}

In the limit of the noncommutativity, i.e. $\Theta\rightarrow0$,
$\eta\rightarrow0$, the above expression of $\mathcal{Z}^{nc}$ tends
to the result of $\mathcal{Z}$ in the usual commutative phase-space,
which is 
\begin{equation}
\mathcal{Z}=\frac{\left(2V\right)^{N}}{\varLambda^{3N}N!}\text{cosh}{}^{N}\left(\beta\mu_{B}B\right).\label{eq:69}
\end{equation}

Using formulae (\ref{eq:37}) and (\ref{eq:38}), we find the magnetization
in noncommutative and commutative phase-space, thus 
\begin{equation}
F^{nc}=-\frac{N}{\beta}\text{ln}\frac{\left(2V\right)}{\varLambda^{3}}\frac{\left(1-\frac{7\Theta\eta}{8\hbar^{2}}\right)\text{cosh}\left(\beta\mu_{B}B\right)}{\left(1+\frac{eB\Theta}{8c\hbar}\right)^{2}}+\frac{1}{\beta}\text{ln}N!,\label{eq:70}
\end{equation}
and the noncommutative magnetization be
\begin{equation}
\left\langle M^{nc}\right\rangle =-\frac{\partial F^{nc}}{\partial B}=2\frac{N}{\beta}\frac{e\Theta}{\left(8c\hbar+eB\Theta\right)}+N\mu_{B}\text{tanh}\left(\beta\mu_{B}B\right).\label{eq:71}
\end{equation}

The commutative magnetization is
\begin{equation}
\left\langle M\right\rangle =-\frac{\partial F}{\partial B}=N\mu_{B}\text{tanh}\left(\beta\mu_{B}B\right),\label{eq:72}
\end{equation}
it is obvious that $\left.\left\langle M^{nc}\right\rangle \right|_{\Theta=0}=\left\langle M\right\rangle $.
We may derive the magnetic susceptibility of electrons $\chi=\frac{1}{V}\frac{\partial\left\langle M\right\rangle }{\partial B}$
in noncommutative phase-space using the magnetization (\ref{eq:71})
by 
\begin{equation}
\chi^{nc}=-2\frac{N}{V\beta}\frac{\left(e\Theta\right)^{2}}{\left(8c\hbar+eB\Theta\right)^{2}}+\frac{N}{V}\beta\mu_{B}^{2}\left(1-\text{tanh}^{2}\left(\beta\mu_{B}B\right)\right),\label{eq:73}
\end{equation}
where the commutative magnetic susceptibility $\chi=\chi^{nc}\left(\Theta=0\right)$
is 
\begin{equation}
\chi=\frac{N}{V}\beta\mu_{B}^{2}\left(1-\text{tanh}^{2}\left(\beta\mu_{B}B\right)\right).\label{eq:74}
\end{equation}

Finally, we conclude with the following special cases. Let us first
consider $B=0$, then we have 
\begin{equation}
\left\langle M\right\rangle =0;\;\left\langle M^{nc}\right\rangle =2\frac{N}{\beta}\frac{e\Theta}{8c\hbar};\text{ and }\;\chi^{nc}=-2\frac{N}{V\beta}\frac{\left(e\Theta\right)^{2}}{\left(8c\hbar\right)^{2}}+\frac{N}{V}\beta\mu_{B}^{2}.\label{eq:76}
\end{equation}

For $B\rightarrow\infty$ and $T=C^{st}$, $\text{tanh}\left(\beta\mu_{B}B\right)=1$,
we obtain 
\begin{equation}
\left\langle M^{nc}\right\rangle =\left\langle M\right\rangle =N\mu_{B};\text{ and }\chi^{nc}=\chi\sim0.\label{eq:77}
\end{equation}

As well when $T\rightarrow\infty$,$\beta\rightarrow0$ (with $B$
be constant), $\text{tanh}\left(\beta\mu_{B}B\right)=0$, we obtain
\begin{equation}
\left\langle M^{nc}\right\rangle \rightarrow\infty,\left\langle M\right\rangle =0;\text{ and }\chi^{nc}\rightarrow\infty,\chi=0.\label{eq:78}
\end{equation}

Armed with the partition function $\mathcal{Z}$, we can compute other
important thermal quantities, such as the average energy $U=-\frac{\partial}{\partial\beta}\text{ln}\mathcal{Z}$,
the entropy $S=\text{ln}\mathcal{Z}-\beta\frac{\partial}{\partial\beta}\text{ln}\mathcal{Z}$
and the specific heat $C=\beta^{2}\frac{\partial^{2}}{\partial^{2}\beta}\text{ln}\mathcal{Z}$.

\section{Conclusion}

In this work, we have exactly studied the three-dimensional Pauli
equation and the corresponding continuity equation for a spin-1/2
particle in the presence of an electromagnetic field in noncommutative
phase-space, considering constant and non-constant magnetic fields.
It is shown that the non-constant magnetic field lifts the order of
the noncommutativity parameter in both the Pauli equation and the
corresponding continuity equation. Given the known absence of the
magnetization current term in the continuity equation, even in noncommutative
phase-space as confirmed by our calculations, we extracted the magnetization
current term from the Pauli equation itself without modifying the
continuity equation. Furthermore, we found that the density current
is conserved, which means, we have a conservation of the deformed
continuity equation.

By using the classical treatment (within the classical limit), the
magnetization and the magnetic susceptibility quantities are explicitly
determined in both commutative and noncommutative phase-spaces through
a semi-classical partition function of the Pauli system of the one-particle
and N-particle systems in three dimensions. Besides, to see the behaviour
of these deformed quantities, we carried out some special cases in
commutative and noncommutative phase-spaces.

Finally, we can say that we successfully examined the influence of
the noncommutativity on the problems in question, where the noncommutativity
was introduced using both the three-dimensional Bopp-Shift transformation
and the Moyal-Weyl product. Further, the noncommutative corrections
to the nonrelativistic Pauli equation and the continuity equation
are also valid up to all orders in the noncommutative parameter. Our
results limits are in good agreement with those obtained by other
authors as discussed and in the literature. 
\begin{acknowledgments}
The author would like to thank Dr. Mojtaba Najafizadeh for his valuable
discussion on the classical partition function.\end{acknowledgments}


\begin{thebibliography}{10}
\bibitem{key-1}Greiner, W. Quantum Mechanics: An Introduction, 4th
ed.; Springer: Berlin, 2001, pp. 339\textendash 340. 

\bibitem{key-2}Ikenberry, E. Quantum Mechanics for Mathematicians
and Physicists. Oxford, New York 1962, pp. 241\textendash 242. 

\bibitem{key-3}Galindo, Alberto, and C. Sanchez del Rio. Intrinsic
magnetic moment as a non-relativistic phenomenon. Am. J. Phys. 1961,
29, 582\textendash 584. \url{https://doi.org/10.1119/1.1937856}

\bibitem{key-4}Nowakowski, M. The quantum mechanical current of the
Pauli equation. Am. J. Phys. 1999, 67 (10), 916-919. \url{https://doi.org/10.1119/1.19149}

\bibitem{key-5}Parker, G. W. Spin current density and the hyperfine
interaction in hydrogen. Am. J. Phys. 1984, 52 (1), 36\textendash 39.
\url{https://doi.org/10.1119/1.13846}

\bibitem{key-6}Shikakhwa, M. S., S. Turgut, and N. K. Pak.. Derivation
of the magnetization current from the non-relativistic Pauli equation:
A comment on \textquoteleft The quantum mechanical current of the
Pauli equation\textquoteright{} by Marek Nowakowski {[}Am. J. Phys.
67 (10), 916 (1999){]}. Am. J. Phys. 2011, 79 (11), 1177\textendash 1179.
\url{https://doi.org/10.1119/1.3630931}

\bibitem{key-7}Hodge, W. B., Migirditch, S. V, and Kerr, William
C. Electron spin and probability current density in quantum mechanics.
Am. J. Phys. 2014, 82 (7), 681\textendash 690. \url{https://doi.org/10.1119/1.4868094}

\bibitem{key-8}Sakurai, J. J. Advanced Quantum Mechanics, (AddisonWesley,
Reading, MA, 1967), pp. 78\textendash 80

\bibitem{key-9}Haouam, I.; Chetouani, L. The Foldy-Wouthuysen transformation
of the Dirac equation in noncommutative Phase-Space. J. Mod. Phys.
2018, 9, 2021\textendash 2034. \url{https://doi.org/10.4236/jmp.2018.911127 }

\bibitem{key-10}Haouam, I. The Phase-Space noncommutativity effect
on the large and small wave-function components approach at Dirac
equation. Open Access Library J. 2018, 5, e4108. \url{https://doi.org/10.4236/oalib.1104108 }

\bibitem{key-11}Bjorken, J.D.; Drell, S.D. Relativistic Quantum Mechanics;
McGraw-Hill, New York, 1964, pp. 10\textendash 13

\bibitem{key-12}Foldy, L.L.; Wouthuysen, S.A. On the Dirac theory
of Spin 1/2 particles and its non-relativistic limit. Phys. Rev. 1950,
78, 29. \url{https://doi.org/10.1103/PhysRev.78.29}

\bibitem{key-13}Pauli, W. Zur Quantenmechanik des magnetischen Elektrons.
Z. Physik 1927, 43, 601\textendash 623. \url{https://doi.org/10.1007/BF01397326}

\bibitem{key-14}Goudsmit. S; Uhlenbeck, George. E. Opmerking over
de Spectra van Waterstof en Helium. Physica 1925, 5, 266\textendash 270. 

\bibitem{key-15}Uhlenbeck, George. E; Goudsmit. S. Ersetzung der
Hypothese vom unmechanischen Zwang durch eine Forderung bezüglich
des inneren Verhaltens jedes einzelnen Elektrons. Die Naturwissenschaften
1925, 13(47), 953\textendash 954. \url{https://doi.org/10.1007/BF01558878}

\bibitem{key-16}Dirac, Paul A. M. The Quantum Theory of the Electron.
Proc. R. Soc. London, Ser. A 1928, 117 (778), 610\textendash 624.
\url{https://doi.org/10.1098/rspa.1928.0023}

\bibitem{key-17}Connes, A. Noncommutative differential geometry.
Publications Mathématiques de L\textquoteright Institut des Hautes
Scientifiques. 62, (1985) 41\textendash 144. \url{https://doi.org/10.1007/BF02698807}

\bibitem{key-18}S.L. Woronowicz. Twisted SU(2) Group. An Example
of a Noncommutative Differential Calculus. Publ. Res. Inst. Math.
Sci. 23, (1987) 117. \url{https://doi.org/10.2977/prims/1195176848}

\bibitem{key-19}Connes, A et al. Noncommutative geometry and Matrix
theory. JHEP 02 (1998) 003. \url{https://doi.org/10.1088/1126-6708/1998/02/003}

\bibitem{key-20}M. M. Sheikh-Jabbari. C, P, and T Invariance of Noncommutative
Gauge Theories. Phys. Rev. Lett. 84 (2000) 5265. \url{https://doi.org/10.1103/PhysRevLett.84.5265}

\bibitem{key-21}Di Grezia, Elisabetta, et al. Spacetime noncommutativity
and antisymmetric tensor dynamics in the early Universe. Phys. Rev.
D 68.10 (2003): 105012. \url{https://doi.org/10.1103/PhysRevD.68.105012}

\bibitem{key-22}O. Bertolami, J. G. Rosa, C. M. L. de Aragão, P.
Castorina, and D. Zappalà. Noncommutative gravitational quantum well.
Phys. Rev. D 72 (2005) 025010. \url{https://doi.org/10.1103/PhysRevD.72.025010 }

\bibitem{key-23}Das, Ashok, et al. Noncommutative supersymmetric
quantum mechanics. Phys. Lett. B 670.4-5 (2009): 407-415. \url{https://doi.org/10.1016/j.physletb.2008.11.011}

\bibitem{key-24}Gracia-Bondia. J. M. Notes on quantum gravity and
noncommutative geometry: New Paths Towards Quantum Gravity. Springer,
Berlin, Heidelberg, 2010. 3-58. \url{https://doi.org/10.1007/978-3-642-11897-5_1}

\bibitem{key-25}Martinetti, P. Beyond the Standard Model with noncommutative
geometry, strolling towards quantum gravity. In Journal of Physics:
Conference Series. IOP Publishing 2015, 634(1), 012001. \url{https://doi.org/10.1088/1742-6596/634/1/012001}

\bibitem{key-26}Seiberg, N; Witten, E. String theory and noncommutative
geometry. J. High Energy Phys. 1999, 9. \url{https://doi.org/10.1088/1126-6708/1999/09/032}

\bibitem{key-27}Christian, K. Quantum groups. No. 155 in Graduate
texts in mathematics, 155. Springer-Verlag, New York, 1995. 

\bibitem{key-28}Connes, Alain; Kreimer, D. Renormalization in quantum
field theory and the Riemann\textendash Hilbert problem I: The Hopf
algebra structure of graphs and the main theorem.Comm Math Phys 2000,
210(1), 249\textendash 273. \url{https://doi.org/10.1007/s002200050779}

\bibitem{key-29}Connes, A; Kreimer, D. Renormalization in Quantum
Field Theory and the Riemann\textendash Hilbert Problem II: The $\beta$-Function,
Diffeomorphisms and the Renormalization Group. Commun. Math. Phys.
2001, 216, 215\textendash 241. \url{https://doi.org/10.1007/PL00005547}

\bibitem{key-30}Tanasa Adrian, Vignes-Tourneret Fabien: Hopf algebra
of non-commutative field theory. J. Noncommut. Geom. 2 (2008), 125-139.
\url{https://doi.org/10.4171/JNCG/17}

\bibitem{key-31}Carroll, S. M., Harvey, J. A., Kostelecký, V. A.,
Lane, C. D., \& Okamoto, T. Noncommutative field theory and Lorentz
violation. Phys. Rev. Lett. 2001, 87(14), 141601. \url{https://doi.org/10.1103/PhysRevLett.87.141601}

\bibitem{key-32}Szabo, R. J. Quantum field theory on noncommutative
spaces. Phys. Rep. 2003, 378(4), 207\textendash 299. \url{https://doi.org/10.1016/S0370-1573(03)00059-0}

\bibitem{key-33}Haouam, I. On the Fisk\textendash Tait equation for
spin-3/2 fermions interacting with an external magnetic field in noncommutative
space-time. J. Phys. Stud. 2020, 24, 1801.\url{https://doi.org/10.30970/jps.24.1801} 

\bibitem{key-34}Kag, L;  Jianhua, W; Chiyi, C. Representation of
noncommutative phase space. Mod. Phys. lett A. 2005, 20, 2165. \url{https://doi.org/10.1142/S0217732305017421}

\bibitem{key-35}Haouam, I. Analytical solution of (2+ 1) dimensional
Dirac equation in time-dependent noncommutative phase-space. Acta.
polytech. 2020, 60(2), 111\textendash 121. \url{https://doi.org/10.14311/AP.2020.60.0111}

\bibitem{key-36}Haouam, I. On the noncommutative geometry in quantum
mechanics. J. Phys. Stud. 2020, 24(2), 2002. \url{https://doi.org/10.30970/jps.24.2002}

\bibitem{key-37}Haouam, I. The Non-Relativistic Limit of the DKP
Equation in Non-Commutative Phase-Space. Symmetry, 2019, 11, 223.
\url{https://doi.org/10.3390/sym11020223}

\bibitem{key-38}Haouam, I. (2019) Continuity Equation in Presence
of a Non-Local Potential in Non-Commutative Phase-Space. Open J. Microphys,
2019, 9, 15-28. \url{https://doi.org/10.4236/ojm.2019.93003}

\bibitem{key-39}Wilkes, J. M. The Pauli and Lévy-Leblond equations,
and the spin current density. Eur. J. Phys. 2020, 41(3), 035402. \url{https://doi.org/10.1088/1361-6404/ab7495}

\bibitem{key-40}Peskin, M. E; Schroeder, D. V. An Introduction to
Quantum Field Theory. Addison-Wesley, New York, 1995, p. 17.

\bibitem{key-41}Chaichian, M., Sheikh-Jabbari, M. M; Tureanu, A.
Hydrogen atom spectrum and the Lamb shift in noncommutative QED. Phys.
Rev. Lett. 2001, 86(13), 2716. \url{https://doi.org/10.1103/PhysRevLett.86.2716}

\bibitem{key-42}Najafizadeh, M; Mehdi, S . Thermodynamics of classical
systems on noncommutative phase space. Chin. J . Phys. 2013, 51(1),
94. \url{https://doi.org/10.6122/CJP.51.94}

\bibitem{key-43}Wei G, F et al. Classical mechanics in non-commutative
phase space$*$. Chin. Phys. C. 2008, 32 338. \url{https://doi.org/10.1088/1674-1137/32/5/002}

\bibitem{key-44}A.E.F. Djemaï and H. Smail. On Quantum Mechanics
on Noncommutative Quantum Phase Spac. Commun. Theor. Phys. 41 (2004)
837. \url{https://doi.org/10.1088/0253-6102/41/6/837}

\bibitem{key-45}Biswas, S. Bohr\textendash van Leeuwen theorem in
non-commutative space. Phys. Lett. A. 2017, 381(44), 3723\textendash 3725.
\url{https://doi.org/10.1016/j.physleta.2017.10.003}\end{thebibliography}
\end{document}